 \documentclass[twocolumn,10pt]{jmlr} 

\usepackage{booktabs}
\usepackage{bm}
\usepackage{makecell}
\usepackage{bbm}
\usepackage[load-configurations=version-1]{siunitx} 

\newcommand{\simil}[1]{\text{sim}\left(#1\right)}
\renewcommand{\exp}[1]{\text{exp}\left(#1\right)}

\newcommand{\equal}[1]{{\hypersetup{linkcolor=black}\thanks{#1}}}

\theorembodyfont{\upshape}
\theoremheaderfont{\scshape}
\theorempostheader{:}
\theoremsep{\newline}

\title[3KG]{3KG: Contrastive Learning of 12-Lead Electrocardiograms using Physiologically-Inspired Augmentations}

\author{%
   \Name{Bryan Gopal}\equal{Equal Contribution} \Email{bryang@cs.stanford.edu}\\
   \addr Stanford University\AND
   \Name{Ryan Han}\footnotemark[1] \Email{ryanhan@cs.stanford.edu}\\
   \addr Stanford University\AND
   \Name{Gautham Raghupathi}\footnotemark[1] \Email{gautham@cs.stanford.edu}\\
   \addr Stanford University\AND
   \Name{Andrew Ng} \Email{ang@cs.stanford.edu}\\
   \addr Stanford University\AND
   \Name{Geoff Tison}\equal{Equal Contribution} \Email{geoff.tison@ucsf.edu}\\
   \addr University of California, San Francisco\AND
   \Name{Pranav Rajpurkar}\footnotemark[2] \Email{pranav\_rajpurkar@hms.harvard.edu}\\
   \addr Harvard Medical School
 }

\begin{document}
\maketitle

\begin{abstract}
We propose 3KG, a physiologically-inspired contrastive learning approach that generates views using 3D augmentations of the 12-lead electrocardiogram. We evaluate representation quality by fine-tuning a linear layer for the downstream task of 23-class diagnosis on the PhysioNet 2020 challenge training data and find that 3KG achieves a $9.1\%$ increase in mean AUC over the best self-supervised baseline when trained on $1\%$ of labeled data. Our empirical analysis shows that combining spatial and temporal augmentations produces the strongest representations. In addition, we investigate the effect of this physiologically-inspired pretraining on downstream performance on different disease subgroups and find that 3KG makes the greatest gains for conduction and rhythm abnormalities. Our method allows for flexibility in incorporating other self-supervised strategies and highlights the potential for similar modality-specific augmentations for other biomedical signals.
\end{abstract}

\begin{center}
\begin{figure*}[h!]
\includegraphics[width=\textwidth]{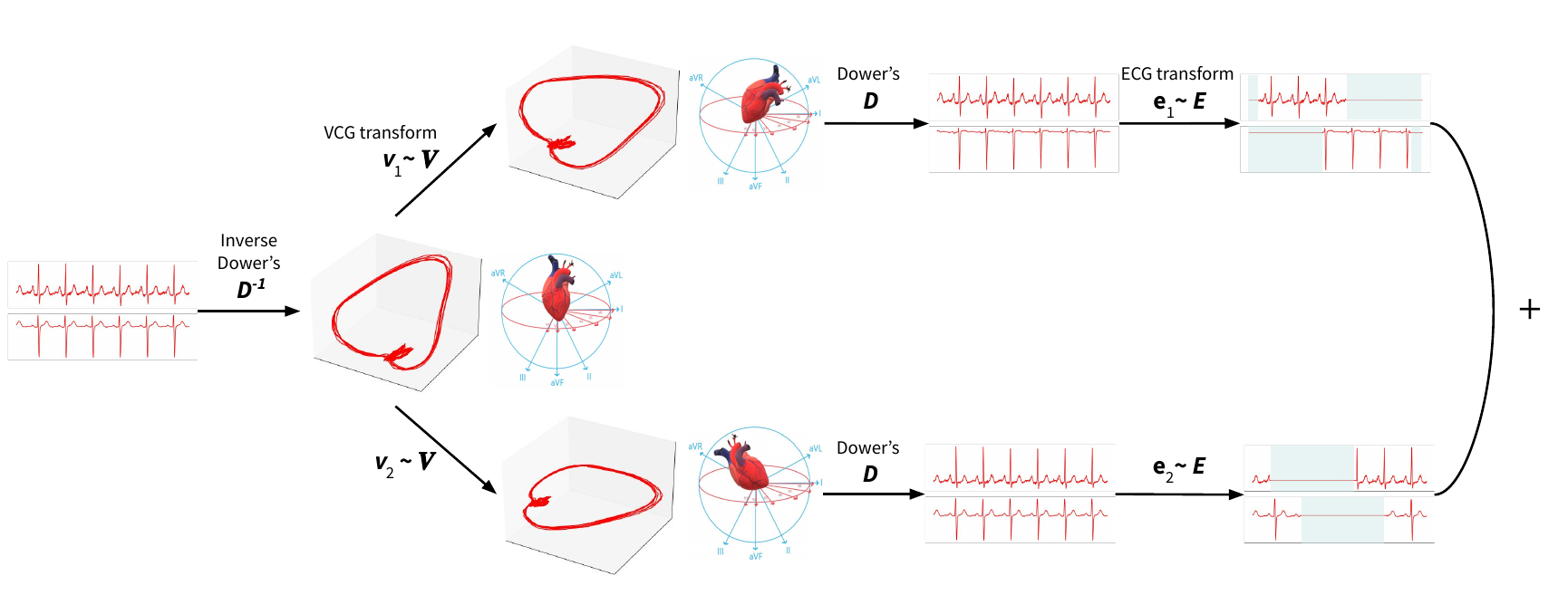}
\caption{\textbf{3KG Overview.} Positive pairs for contrastive learning are generated by composing stochastic augmentations of the 12-lead ECG in the VCG and ECG spaces respectively. Leads are encoded independently and then treated as positive views if they derive from the same patient. Only leads II and V2 are shown in the figure for the sake of simplicity.}
\label{fig:teaser}
\end{figure*}
\end{center}

\section{Introduction}

The 12-lead electrocardiogram (ECG) is a common, non-invasive test for the screening, diagnosis, and management of cardiovascular conditions \citep{salerno2003competency, fesmire1998usefulness}. Recent studies have demonstrated the ability of deep learning to predict conditions commonly diagnosed by physicians on the ECG as well as novel diagnoses beyond those traditionally identifiable by physicians with an ECG \citep{tison2019automated, attia2019artificial, attia2019screening}. However, such supervised methods depend on tens of thousands of high quality labels to achieve strong generalization performance: we find that supervised learning is only marginally better than random for certain diagnoses at low label fractions. There is therefore a need for label-efficient learning strategies to reduce the resources required for algorithmic training. 

Self-supervised contrastive learning presents a promising avenue for improving the label-efficiency of ECG interpretation strategies. Contrastive learning approaches leverage unlabeled data to pretrain models for fine-tuning on downstream tasks where labeled data is scarce \citep{hjelm2019learning, wu2018unsupervised}. On computer vision tasks, contrastive learning approaches have reached performance comparable to that of of supervised learning while using as little as $1\%$ of the labeled data \citep{chen2020simple, he2020momentum, vu2021medaug}. However, there still remains a large performance disparity between supervised learning and contrastive methods for ECG analysis that are fine tuned on small label fractions \citep{kiyasseh2020clocs, cheng2020subjectaware, mehari2021ssl, diamant2021pcl}.

In this work, we propose 3KG, a physiologically-inspired contrastive learning approach that leverages the unique spatiotemporal properties of the ECG. Our method generates views for contrastive learning using ECG-specific 3D augmentations in the vectorcardiogram (VCG) space.
VCGs derive from ECGs and represent the electrical activity of the heart along three orthogonal spatial axes. We hypothesize that VCGs enable spatial manipulations of ECGs in a manner consistent with the underlying physiology they represent. After spatial manipulation, we project the VCG back in to the ECG space and apply temporal manipulations. We further extend this method by incorporating the spatial relationship between the leads by encoding them separately and treating them as positive views. An overview for this approach is provided in Figure \ref{fig:teaser}. 

We conduct pretraining and evaluate the quality of learned representations on the PhysioNet 2020 dataset \citep{PhysioNet2020} for the task of arrhythmia classification. Our contributions are as follows:

\begin{enumerate}
    \item We demonstrate that 3KG outperforms previous self-supervised ECG strategies and closes the gap to achieving fully supervised performance when only using a subset of the labeled data. Our best performing model achieves an AUROC of 0.826 when trained on $1\%$ of labeled data compared to an AUROC of 0.898 for fully supervised learning on $100\%$ of labeled data. For reference, we find that the previous best self-supervised learning approach achieves an AUROC of 0.757 when trained on the $1\%$ label fraction.
    \item We perform comparative empirical analyses with spatial augmentation strength and find that stronger spatial augmentations lead to better learned representations. Additionally, we observe that combining spatial augmentations increases performance over any individual spatial augmentation, and that combining spatial augmentations with temporal augmentations produces the strongest results of all.
    \item  We investigate how the introduction of an inductive bias through use of physiologically-inspired pretraining affects downstream performance on different disease subgroupings relative to a randomly initialized baseline. We find that 3KG makes the greatest gains for conduction and rhythm abnormalities and smaller gains for waveform abnormalities.
\end{enumerate}

\section{Related Work}

Many contrastive approaches have been developed for computer vision tasks \citep{chen2020simple, he2020momentum, grill2020bootstrap}. These methods define positive views, the representations of which are pushed together during pretraining, and negative views, the representations of which are pulled apart \citep{tian2020makes}. The definition of views thus controls the semantic information encoded by the learned representations. Examples of definitions include \cite{tian2019multiview} who use different modal views of the same image as positive views, and \cite{chen2020simple} who define positive views as differently augmented views of the same image. \cite{chen2020simple} note that augmentation strength and composition are crucial for learning good representations; the development of augmentations applicable to ECGs is thus important.

Self-supervised methods have also been applied to the ECG domain. While \cite{sarkar2020emotion} pretrain their model on a pretext task of predicting which of six signal transformations (gaussian noise, scaling, temporal inversion, negation, permutation, and time-warping) was applied, \cite{mehari2021ssl} adapt SimCLR to the ECG domain through the use of random resized crops and time masking as augmentations. They additionally apply Contrastive Predictive Coding \citep{oord2018representation} to ECGs. Although these methods improve over randomly initialized baselines, they do not take advantage of the unique properties of the ECG as they are largely direct extensions from other domains.

Other contrastive ECG methods incorporate unique attributes of the ECG. \cite{mehari2021ssl} propose the use of physiological noise (baseline wander, powerline noise, electromyographic noise, and baseline shift) as augmentations for SimCLR. In Patient Contrastive Learning, \cite{diamant2021pcl} define positive pairs as different ECGs taken from the same patient at varying points in time. In contrast to our approach, these methods emphasize the clinical nature of the ECG recording without taking advantage of its spatiotemporal attributes. Another approach by \cite{kiyasseh2020clocs} proposes a trio of methods that leverage the spatiotemporal nature of the ECG by treating different segments, leads, and both segments and leads, as positive views. However, this approach limits itself to the ECG space, while our approach takes advantage of natural augmentations to the ECG in the VCG space.

\section{Methods}




\subsection{Problem Formulation}
Given a batch of $N$ 12-lead ECGs $\bm{x}$, a set of positive view pairs $\mathcal{P}$, a space of transformations $\mathcal{T}$, and an additional positive view condition $\mathcal{C}$, our goal during pretraining is to minimize a contrastive loss function

\begin{align*}
\mathcal{L}(\bm{x}) &= \frac{1}{|\mathcal{P}|}\sum_{i, j \in \mathcal{P}} \ell_{i, j}(\bm{x})\\
\ell_{i, j}(\bm{x}) &= -\log \frac{s_{i, j}(\bm{x})}{\sum_{k=1}^{2N} \mathbbm{1}_{\left[k \neq i\right]} \mathbbm{1}_{\left[i, k \not \in \mathcal{P} \right]} \mathbbm{1}_{\neg\mathcal{C}(i, k)}\text{ }s_{i, k}(\bm{x})}\\
s_{i, j}(\bm{x}) &= \exp{\simil{\bm{z}_i, \bm{z}_j} / \tau}\\
\end{align*}

where the $L_2$-normalized output embedding $\bm{z}_i$ is produced by passing ECG sample $\bm{x}_i$ through our pretraining model after it has been augmented with a transformation sampled from $\mathcal{T}$. The function $\text{sim}$ denotes the cosine similarity function, and $\tau$ is a temperature hyperparameter.

Following \cite{chen2020simple}, we do not sample negative views explicitly, instead defining a negative view for index $i$ as all examples in the batch that are not part of a positive pair with $i$ and fail to satisfy condition $\mathcal{C}$.

$\mathcal{T}$, $\mathcal{C}$, and $\mathcal{P}$ all contribute to the definition of what a positive view of an ECG sample $x$ can be. We use physiologically-based inductive biases to propose \textit{3KG}, which describes novel formulations for all three.

\subsection{Positive View Definition}
\subsubsection{Augmentations}

We seek to leverage the vectorcardiogram (VCG) in combination with time masking to build a space of transformations $\mathcal{T}$ that preserves spatiotemporal features relevant for assessment of ECG signals.

The vectorcardiogram (VCG) represents the gold standard in visualizing the spatial propagation of a heart signal over time. In contrast to a 12-lead ECG (which provides 12 2D projections of a 3D signal), a VCG shows the electrical activity of the heart along three orthogonal spatial axes, tracing contours around a central point over time.

Since natural variability in cardiac structure and orientation exists without substantive clinical implications, we argue that patient VCGs can be similarly augmented through rotation and scaling without breaking the spatiotemporal invariances needed to make downstream diagnoses.

We also apply time masking on a per-lead basis as a temporal augmentation. Taking inspiration from self-supervised approaches in speech recognition \citep{baevski2020wav2vec}, we choose to implement time masking in order to mimic how ECG assessment is done in a clinical setting. Since cardiologists often only need a single beat on a single lead to ascertain a significant portion of ECG diagnoses, a self-supervised model should similarly be able to learn useful representations of ECGs that have some beats masked.

We formalize our transformation space as follows:
\[\mathcal{T} = \mathcal{E} \circ \mathcal{V},\] 

where $\mathcal{E}$ is the space of all temporal ECG augmentations and $\mathcal{V}$ is the space of all spatial vectorcardiogram (VCG) perturbations. We define $\mathcal{V}$ to be

\[\mathcal{V}(\bm{x}_i) = DSRD^{-1}\bm{x}_i,\]

where 
\begin{itemize}
    \item the Inverse Dowers transformation $D^{-1} \in \mathbb{R}^{3 \times 12}$ projects the $12$-lead ECG into $3$-dimensional vectorcardiogram space \citep{InverseDowers}.
    \item a rotation matrix $R \in \mathbb{R}^{3 \times 3}$ performs a 3-dimensional rotation of a VCG along all three orthogonal axes.
    \item a scaling matrix $S \in \mathbb{R}^{3 \times 3}$ scales the vectorcardiogram along all three orthogonal axes.
    \item the Dowers transformation $D \in \mathbb{R}^{12 \times 3}$ projects the VCG back into 12-lead ECG space \citep{InverseDowers}
\end{itemize}
Note that because VCGs exist in a lower-dimensional space than ECGs, the conversion process between the two observes some loss since $D$ approximates an inverse to the underdetermined system $D^{-1}$.

Similarly, we define $\mathcal{E}$ to be the space of all possible time masks for each of the 12 ECG leads. 

In practice, we apply our augmentation scheme in the following order:

\begin{enumerate}
    \item \textbf{VCG Augmentations} $\bm{\mathcal{V}}$
    \begin{enumerate}
        \item \textbf{Convert ECG to VCG.} We apply the Inverse Dowers transformation $D^{-1}$ to compute a VCG from our 12-lead ECG.
        \item \textbf{Random Rotation.} We rotate the VCG on all three axes in a random order, where each per-axis rotation is a random value uniformly sampled between $-r$ to $+r$ degrees for a user-specified value of $r$.
        \item \textbf{Random Scale.} We scale the VCG on all three axes, where each per-axis scale is a random value uniformly sampled from a uniform distribution between $[1,s]$ with a $50\%$ probability of inversion for a user-specified value of $s$. This means that on average, one half of the scale factors is between $[1/s, 1]$ and the other half is between $[1, s]$.
        \item \textbf{Convert VCG to ECG.} We apply the Dowers transformation $D$ to compute an ECG from our VCG.
    \end{enumerate}
    \item \textbf{ECG Augmentations} $\bm{\mathcal{E}}$
    \begin{enumerate}
        \item \textbf{Random Time Masking.}  \label{tmask} We uniformly sample 1 time step for each lead as a starting index, and replace the subsequent $\lfloor L \cdot p_t \rfloor$ time steps to be zero over all leads (wrapping around to the beginning of the recording as necessary), where $L$ is the number of time steps in the sample and $0 \leq p_t \leq 1$ is the user-specified time mask percentage.
    \end{enumerate}
\end{enumerate}

We sample two transformations $t_1, t_2 \sim \mathcal{T}$ at each training step and apply each to the batch, resulting in the positive pair $[t_1(\bm{x}_i), t_2(\bm{x}_i)]$ for each 12-lead ECG $\bm{x}_i$ (see \autoref{fig:teaser} for more details).

\subsubsection{Patient-Aware Views}
We seek to ensure that ECG recordings that belong to the same patient are not considered as negative views of each other. We thus couple a recording number label to each ECG sample during pretraining and let our additional condition $\mathcal{C}$ be "ECG samples that originate from the same recording are considered positive views of each other". This additional condition transforms our traditional contrastive loss function into a \textit{Patient-Supervised Contrastive Loss} and is in line with previous works that have also taken a patient-aware approach in loss calculation \citep{kiyasseh2020clocs, diamant2021pcl}.



\subsubsection{Per-Lead Views} \label{simcg}
We seek to leverage the inherent spatial relationship between individual ECG leads to learn better representations during pretraining. Some previous contrastive methods for 12-lead ECGs have treated different leads as independent channels akin to color channels in computer vision \citep{diamant2021pcl, mehari2021ssl}. Because the 12 leads in an ECG provide different spatial views of the same underlying cardiac electrical activity, they can also be thought of as natural augmentations of each other \citep{kiyasseh2020clocs}.

We thus propose modifying our set of positive pairs $\mathcal{P}$ to consider individual leads as positive views. We evaluate the impact of this formulation by comparing models trained with and without per-lead views, keeping all other parts of experimental setup the same. We denote our two different methods for handling the 24 lead embeddings produced per positive pair $[t_1(\bm{x}_i), t_2(\bm{x}_i)]$ as follows:
\begin{itemize}
    \item \textbf{SimCG-A}, which \textit{does not} use a per-lead formulation of $\mathcal{P}$ and instead averages the 12 embeddings produced per ECG. As such, there is only positive embedding pair per ECG sample $\bm{x}_i$, with each of the two component embeddings coming from $t_1(\bm{x}_i)$ and $t_2(\bm{x}_i)$.
    \item \textbf{SimCG-L}, which \textit{does} use a per-lead formulation of $\mathcal{P}$, constructing positive pairs over all lead embeddings during loss computation. As such, we arrive at $\binom{24}{2} = 276$ positive embedding pairs per ECG $\bm{x}_i$.
\end{itemize}

We note that SimCG-L is similar to CMLC \citep{kiyasseh2020clocs} in that different leads from the same patient are treated as positive pairs. However, notable differences include the fact that our approach incorporates spatiotemporal augmentations and uses a nonlinear projection head during pretraining to improve the representation quality of the final layer of the encoder. Additionally, we accumulate loss across all pairs of views of an ECG, which we did not find implemented in the released codebase.

Finally, we denote \textbf{3KG} to be the combination of our spatiotemporal augmentation strategy, Patient-Supervised Contrastive Loss, and SimCG-L.

\begin{table*}[!ht]
\centering
\begin{tabular}{l|rrr|rrr}
\hline
\multicolumn{1}{c|}{Augmentation} & \multicolumn{3}{c|}{AUROC}                                                         & \multicolumn{3}{c}{F1}                                                        \\
\multicolumn{1}{c|}{}       & \multicolumn{1}{c}{1\%}   & \multicolumn{1}{c}{10\%}  & \multicolumn{1}{c|}{100\%} & \multicolumn{1}{c}{1\%} & \multicolumn{1}{c}{10\%} & \multicolumn{1}{c}{100\%} \\ \hline
\textit{Maximum Rotation (degrees)}    & \multicolumn{1}{l}{}      & \multicolumn{1}{l}{}      & \multicolumn{1}{l|}{}      & \multicolumn{1}{l}{}    & \multicolumn{1}{l}{}     & \multicolumn{1}{l}{}      \\
5                           & 0.771                    & 0.835                     & 0.851                     & 0.286                   & 0.366                    & 0.395                     \\
10                          & 0.783                    & 0.839                    & 0.858                     & 0.295                   & 0.374                    & 0.404                     \\
25                          & 0.800                    & 0.854                    & 0.873                     & 0.312                   & 0.386                    & 0.418                     \\
45                          & \textbf{0.801}           & \textbf{0.856}           & \textbf{0.877}            & \textbf{0.315}          & \textbf{0.386}           & \textbf{0.422}            \\ \hline
\textit{Maximum Scale Factor}       & \multicolumn{1}{l}{}      & \multicolumn{1}{l}{}      & \multicolumn{1}{l|}{}      & \multicolumn{1}{l}{}    & \multicolumn{1}{l}{}     & \multicolumn{1}{l}{}      \\
1.05                        & 0.771                    & 0.831                    & 0.849                     & 0.273                   & 0.360                    & 0.392                     \\
1.1                         & 0.764                    & 0.834                     & 0.851                     & 0.260                   & 0.366                    & 0.398                     \\
1.5                         & \textbf{0.795}           & \textbf{0.852}           & \textbf{0.868}            & \textbf{0.303}          & \textbf{0.379}           & \textbf{0.408}            \\
2                           & 0.789                    & 0.846                    & 0.866                     & 0.299                   & 0.381                    & 0.412                     \\ \hline
\textit{Combinations}       & \multicolumn{1}{l}{}      & \multicolumn{1}{l}{}      & \multicolumn{1}{l|}{}      & \multicolumn{1}{l}{}    & \multicolumn{1}{l}{}     & \multicolumn{1}{l}{}      \\
Rotate 45 + Time Mask             & \textbf{0.812}            & \textbf{0.870}            & 0.881                      & 0.306                   & 0.397                    & 0.428                     \\
Scale 1.5 + Time Mask             & 0.798                     & 0.859                     & 0.878                      & 0.302                   & 0.387                    & 0.417                     \\
Rotate 45 + Scale 1.5             & 0.807                     & 0.861                     & 0.881                      & 0.319                   & 0.394                    & 0.431                     \\
Rotate 45 + Scale 1.5 + Time Mask & \textbf{0.812}                     & 0.869                     & \textbf{0.883}             & \textbf{0.320}          & \textbf{0.400}           & \textbf{0.432}            \\ \hline
\end{tabular}
\caption{\textbf{Linear Evaluation for Different Augmentation Combinations on SimCG-A.} The time mask percentage was set to 50\% for all experiments. Combinations were composed of best individual augmentations.}
\label{table:augmentations}
\end{table*}

\subsection{Experimental Setup}
\subsubsection{Dataset}
We run all experiments on the PhysioNet 2020 challenge training data \citep{PhysioNet2020}. The dataset consists of 12-lead ECG recordings from four data sources with various distributions of patient characteristics, recording length, and diagnoses. There are a total of 43,134 recordings which we split into training, validation, and test sets in an 80/10/10 configuration; there was no patient overlap between these sets. The training set was further divided into 1\%, 10\%, and 100\% label fraction splits to simulate label scarcity for the downstream task. Labels are at the recording level, and each recording could be associated with multiple labels. Although the training data had 111 classes present, we chose to classify only the 27 SNOMED CT codes that were included in the challenge evaluation metric. A board-certified cardiologist coalesced the following SNOMED CT labels after the determination that the diagnoses were substantially equivalent:  complete right bundle branch block and right bundle branch block, premature atrial contraction and supraventricular premature beats, premature ventricular contractions and ventricular premature beats, 1st degree atrioventricular block and prolonged PR interval. Our final downstream task was a 23-class multilabel classification.

\paragraph{Preprocessing.}

To ensure ECGs from different sources were standardized, all recordings were resampled to 500 Hz and then exhaustively cropped into disjoint 5-second segments. All crops were tagged with the file they originated from for later use in loss computation. ECGs were normalized to unit range, but were not translated so as to maintain the isoelectric line at 0.

\subsubsection{Architectures and Optimization} \label{archs-brief}

Our encoder is a 1-D convolutional neural network with a single channel input and 256-dimensional output (see Appendix \ref{archs-appendix} for more details).  We chose to use 1 input channel and process each lead separately to increase the versatility of our model to settings where not all 12 leads are available, such as in the ambulatory (i.e. Holter) setting where the standard of care is a single lead ECG \citep{KENNEDY1992341}.

Following similar work in self-supervision, we leveraged a 2-layer multilayer perceptron (MLP) network in pretraining that projected this 256-dimensional output into 128-dimensional space \citep{chen2020simple}. We used the Adam optimizer with a learning rate of $1 \times 10^{-3}$ and a batch size of $512$ for all pretraining experiments \citep{kingma2017adam}.

\subsubsection{Evaluation on Downstream Task}
We perform a downstream linear evaluation of our pretrained encoder across 1\%, 10\%, and 100\%  label fractions. After pretraining, we freeze our encoder's parameters and drop our MLP, attaching a single linear layer to predict diagnosis labels from the learned output representations. We average the normalized single lead embeddings generated by passing a 12-lead ECG crop into our single-lead encoder. This final averaged embedding is sent downstream to the linear classifier. The linear classifier is trained using a binary cross-entropy loss averaged over all classes. For the 1\% and 10\% label fractions, we trained 5 different linear classifiers on 5 different fractional splits and used all of their output predictions in metric calculation. We report the AUROC and F1 scores averaged across all 23 classes in the test set for our experimental results. 

We used the Adam optimizer and a batch size of $512$ for all downstream experiments \citep{kingma2017adam}. For each experiment, we performed an automatic learning rate search over the space $\left[10^{-8}, 1\right]$ following the methodology outlined by \cite{smith2017cyclical}. In practice, we found that the optimal learning rate for the linear layer landed within the range $\left[0.05, 0.2\right]$ for most experiments.

\paragraph{Statistical Testing.}
For each experiment, we evaluate the statistical significance of each of our experiments against a stated baseline model using a non-parametric bootstrapping setup. We randomly sample $N_T$ test set predictions with replacement (where $N_T$ is the size of the test set) from both the baseline and the model of interest. We then compute the metric difference between the two resampled prediction sets. We repeat this entire process 1000 times, and construct a 95\% confidence interval using all the resulting differences. If $0$ lies outside this interval, we consider the model of interest to be significantly different than the baseline model.

\begin{center}
\begin{table*}[!ht]
\centering
\resizebox{\textwidth}{!}{
\begin{tabular}{l|rrr|lll}
\hline
\multicolumn{1}{c|}{Method}                                       & \multicolumn{3}{c|}{AUROC}                                                         & \multicolumn{3}{c}{F1}                                                        \\
\multicolumn{1}{c|}{}                                             & \multicolumn{1}{c}{1\%}   & \multicolumn{1}{c}{10\%}  & \multicolumn{1}{c|}{100\%} & \multicolumn{1}{c}{1\%} & \multicolumn{1}{c}{10\%} & \multicolumn{1}{c}{100\%} \\ \hline
\textit{Supervised Baseline} & \multicolumn{1}{l}{}      & \multicolumn{1}{l}{}      & \multicolumn{1}{l|}{}      &                         &                          &                           \\
Random Initialization                                                              & 0.694          & 0.836           & \textbf{0.898}             & 0.170          & \textbf{0.349}           & \textbf{0.444}           \\ \hline
\textit{Self-Supervised Baselines} & \multicolumn{1}{l}{}      & \multicolumn{1}{l}{}      & \multicolumn{1}{l|}{}      &                         &                          &                           \\
Patient Contrastive Learning \citep{diamant2021pcl}                                                              & \textbf{0.757}          & \textbf{0.843}           & 0.857             & \textbf{0.259}          & 0.335           & 0.362           \\
Contrastive Multi-Segment Coding \citep{kiyasseh2020clocs}                                                             & 0.728                   & 0.782                    & 0.792                      & 0.210                   & 0.277                    & 0.308                     \\
Contrastive Multi-Lead Coding \citep{kiyasseh2020clocs}                                                             & 0.730                   & 0.769                    & 0.779                      & 0.205                   & 0.274                    & 0.321                     \\
Contrastive Multi-Segment/Lead Coding \citep{kiyasseh2020clocs}                                                           & 0.711                   & 0.774                    & 0.788                      & 0.201                   & 0.263                    & 0.299                     \\
SimCLR + RRC/TO \citep{mehari2021ssl}                                    & 0.594                   & 0.634                    & 0.740                      & 0.121                   & 0.153                    & 0.242                     \\
SimCLR + Physio. \citep{mehari2021ssl}                                            & 0.608                   & 0.641                    & 0.744                      & 0.141                   & 0.163                    & 0.229                     \\
Contrastive Predictive Coding \citep{mehari2021ssl}                                                               & 0.554                   & 0.573                    & 0.677                      & 0.099                   & 0.093                    & 0.170                     \\
Signal Transformation \citep{sarkar2020emotion}                                               & 0.548                   & 0.562                    & 0.564                      & 0.085                   & 0.085                    & 0.099                     \\
 \hline
\textit{Our Methods}                                              & \multicolumn{1}{l}{}      & \multicolumn{1}{l}{}      & \multicolumn{1}{l|}{}      &                         &                          &                           \\
SimCG-A + VCG + Time Mask                                          & 0.812                   & 0.869                    & 0.883                      & \textbf{0.320}                   & 0.400                    & 0.432                     \\
SimCG-L                                                             & 0.789                   & 0.851                    & 0.869                      & 0.304                   & 0.389                    & 0.420                     \\
3KG (SimCG-L + VCG + Time Mask)                                      & \textbf{0.826}                   & \textbf{0.876}                    & \textbf{0.890}                      & 0.309                   & \textbf{0.411}                    & \textbf{0.438}                     \\ \hline
\end{tabular}
}
\caption{\textbf{PhysioNet2020 performance of linear classifiers trained on representations learned with self-supervised methods.} Augmentation parameters (Rotate 45, Scale 1.5) were chosen based on highest performance on SimCG-A. Differences between the top baseline (Patient Contrastive Learning) and all our methods are statistically significant for all metrics and fractions displayed. The top performer among baselines and top performer among our methods are bolded. Abbreviations: Random Resized Crop (RRC), Timeout (TO), Physiological Noise (Physio).}
\label{table:baselines}
\end{table*}
\end{center}

\section{Experiments}

We use AUROC as our primary metric to evaluate performance across the 23 classes, and also report F1 scores. Our results are shown in Table \ref{table:augmentations} and \ref{table:baselines}.




\subsection{Spatial VCG Augmentations}
\paragraph{Design} We compare the performance of different augmentation parameters and examine the improvement in performance due to a combination of augmentations. \cite{tian2020makes} argue that a good view in contrastive learning retains task-relevant information while reducing the mutual information between views. We are motivated to generate positive views of an ECG signal by applying physiologically-inspired augmentations under this principle. \cite{chen2020big} show that the combination of augmentations in a contrastive learning setup could amount to gains in performance greater than the gains in performance of any individual augmentation. Therefore, we are also motivated to combine our physiologically-inspired augmentations with time masking, an augmentation commonly used for signal data \citep{baevski2020wav2vec}. We choose to evaluate our physiologically-inspired augmentations with SimCG-A, since we want to isolate the impact of our augmentations to better optimize the augmentation parameters. We default to 50\% for the user-specified Time Mask percentage as that is the optimal percentage found by \cite{baevski2020wav2vec}.

\paragraph{Single Augmentation Results.} We find Rotate 45 and Scale 1.5 to be the strongest augmentation parameters (as seen in \autoref{table:augmentations}). Among Rotate 5, 10, 25, and 45, we find that Rotate 45 is our best performer at the 1\% (0.801), 10\% (0.856), and 100\% (0.877) label fractions. Among Scale 1.05, 1.1, 1.5, and 2, we find that Scale 1.5 is our best performer at the 1\% (0.795), 10\% (0.852), and 100\% (0.868) label fractions.

\paragraph{Combination Augmentation Results.} We find Rotate 45 + Scale 1.5 + Time Mask to be the best performing combination overall as seen in Table \ref{table:augmentations}. The combination of Rotate 45 + Scale 1.5 performs better than Rotate 45 or Scale 1.5 alone at the 1\% (0.807 vs 0.801 \& 0.795), 10\% (0.861 vs 0.856 \& 0.852), and 100\% (0.881 vs 0.877 \& 0.868) label fractions. When adding Time Mask, we find that Rotate 45 + Time Mask outperforms Rotate 45 at the 1\% (0.812 vs 0.801), 10\% (0.870 vs 0.856), and 100\% (0.881 vs 0.877) label fractions. Similarly, Scale 1.5 + Time Mask outperforms Scale 1.5 at the 1\% (0.798 vs 0.795), 10\% (0.859 vs 0.852), and 100\% (0.878 vs 0.868) label fractions. Rotate 45 + Scale 1.5 + Time Mask underperforms Rotate 45 + Time Mask at the 10\% (0.869 vs 0.870) label fraction. However, it is still the best performer at the 1\% (0.812) and 100\% (0.883) label fractions.





\begin{center}
\begin{figure*}[h!t]
\includegraphics[width=\textwidth]{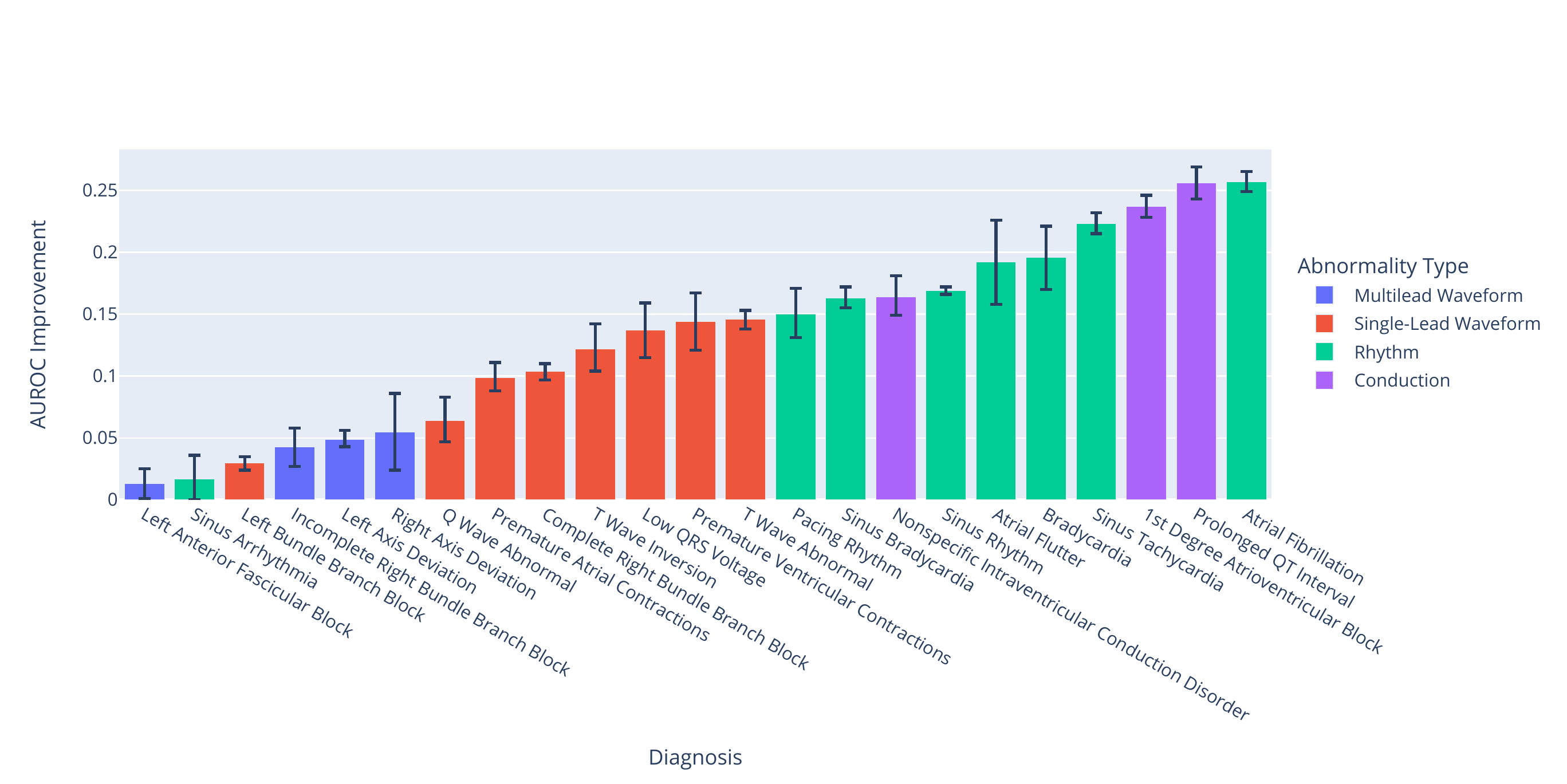}
\caption{\textbf{Per-Diagnosis AUROC Comparisons}. We show the per-diagnosis improvement in AUROC that 3KG (linear evaluation) has over a fully supervised, randomly-initialized model, both trained on 1\% of labels. 95\% confidence intervals for the improvements on each diagnosis are also shown. A board-certified cardiologist classified the 23 abnormalities into four types: multilead waveform, single-lead waveform, rhythm, and conduction. All improvements on all diagnoses are statistically significant.
}
\label{fig:figure3}
\end{figure*}
\end{center}

\subsection{Comparison to Previous Methods}
We compare our best performer, 3KG (SimCG-L + VCG + Time Mask), with 8 prior self-supervised learning methods applied to the ECG domain. 
\paragraph{Reproduction.} To represent previous works as faithfully as possible, we adapt their publicly released codebases to accomodate the PhysioNet 2020 challenge dataset and modify their encoders to use our standardized encoder architecture. The following methods use a 12-channel encoder: Patient Contrastive Learning (PCLR) \citep{diamant2021pcl}, SimCLR + RRC/TO (Random Resized Crop/Timeout) \citep{mehari2021ssl}, SimCLR + Physio. (Physiological Noise) \citep{mehari2021ssl}, and Contrastive Predictive Coding (CPC) \citep{mehari2021ssl}. In contrast, the following methods use a 1-channel encoder: Signal Transformation \citep{sarkar2020emotion}, Contrastive Multi-Segment Coding (CMSC) \citep{kiyasseh2020clocs}, Contrastive Multi-Lead Coding (CMLC) \citep{kiyasseh2020clocs}, and Contrastive Multi-Segment/Lead Coding (CMSMLC) \citep{kiyasseh2020clocs}. We pretrain models using their out of the box training procedures and configurations. The released implementation for PCLR did not include a training pipeline, so we pretrained this baseline in our codebase. We loaded all pretrained models into our unified codebase for downstream evaluation to ensure that output metric results are identical.


\paragraph{Results.} Each of our proposed methods (SimCG-A + VCG + Time Mask, SimCG-L, and 3KG) significantly outperforms all 8 self-supervised baselines, as shown in Table \ref{table:baselines}. Among our baselines, we find PCLR to be the top performer. SimCG-A + VCG + Time Mask and SimCG-L individually outperform PCLR at the 1\% (0.812 \& 0.789 vs 0.757), 10\% (0.869 \& 0.851 vs 0.843), and 100\% (0.883 \& 0.869 vs 0.857) label fractions. 3KG outperforms PCLR by an even greater difference at the 1\% (0.826 vs 0.757), 10\% (0.876 vs 0.843), and 100\% (0.890 vs 0.857) label fractions. This experiment shows the independent value of SimCG-A + VCG + Time Mask and SimCG-L over our baselines, as well as the additional value of combining these methods in 3KG.

\subsection{Per-Pathology Performance Analysis}

We explore 3KG at the per-diagnosis level to better understand how our physiologically-inspired self-supervised learning scheme affects the quality of the learned representations. We compare the linear evaluation of our best performer 3KG to a fully supervised, randomly-initialized model where both are trained on the same 1\% label fraction. A board-certified cardiologist classified the diagnosis into four different abnormality types: multilead waveform, single-lead waveform, rhythm, and conduction. 

\paragraph{Results.} As shown in Figure \ref{fig:figure3}, 3KG generally makes greater AUROC gains for conduction (mean = 0.22) and rhythm (mean = 0.17) abnormalities and smaller AUROC gains for single-lead waveform (mean = 0.12) and multilead waveform (mean = 0.04) abnormalities. All improvements across all diagnoses are statistically significant.
\\
\section{Discussion}

We introduce 3KG, a method to use the spatiotemporal properties of the ECG to generate and select positive views for contrastive learning, and demonstrate the utility of this method on an arrythmia diagnosis dataset.

\paragraph{Can we improve performance by incorporating physiological constraints into our positive pairs?} Yes. Our best pretrained strategy obtains a $19.0\%$ increase in AUROC on the $1\%$ fraction over a supervised baseline, as well as a $9.1\%$ increase over the approach implemented in \cite{diamant2021pcl}, which we found to be the strongest previous self-supervised approach on our dataset.

Our empirical analysis shows that stronger spatial augmentations lead to better learned representations with diminishing returns at higher augmentation strengths. Performance eventually decreases for scaling at the higher end of the range (from 1.5 to 2) but no decrease is observed for rotation, although this may be due to the limited range of sampled strengths. Additionally, we observe that combining spatial augmentations increases performance over any individual spatial augmentation, and that combining spatial augmentations with temporal augmentations produces the strongest results of all. A similar result has been observed by \cite{chen2020simple} in the computer vision domain demonstrating augmentation strength and composition are both crucial for learning good representations.

To the best of our knowledge, we are the first to investigate how the introduction of an inductive bias through contrastive pretraining affects downstream performance on different disease subgroupings. Compared to a randomly initialized baseline, we find that 3KG makes the greatest gains for conduction and rhythm abnormalities and smaller gains for waveform abnormalities. However, it is hard to attribute these results to a specific component of 3KG due to the various components of our best performing model.

\paragraph{Limitations.}
Some limitations of this work should be noted. First, we do not perform an exhaustive hyperparameter search to determine the ideal time mask percentage, rotation, and scale amounts. Second, even though our dataset was comprised of multiple independent sets of ECG recordings collected with different institutions and included additional metadata, we did not explore the utility of this additional metadata in our method \citep{vu2021medaug}. Finally, we did not formally investigate the relationship between encoder size and performance, instead opting to use a relatively small network in both pretraining and fine-tuning due to computing resource constraints.

In closing, our work demonstrates the benefits of incorporating spatiotemporal augmentations into contrastive learning for ECGs and highlights the potential for similar modality-specific augmentations for other biosignals. The introduction of an inductive bias through physiologically-inspired pretraining is especially relevant for clinicians, who may be more inclined to trust algorithms trained with intuitive procedures that reflect well-understood physiological principles. Strengths of our work include the fact that our approach is easily extensible in that our framework allows flexibility in incorporating other self-supervised strategies; additionally, all data used for experiments are publicly available, as will the codebase be that was used to implemented our methods.


\newpage
\bibliography{bibliography}

\newpage
\appendix
\section{Model Architectures}
\label{archs-appendix}
The convolutional neural network we used for our experiments is described below, with all convolutions being 1-dimensional and all activations being Leaky ReLUs with a negative slope of 0.2:
\begin{center}
\resizebox{\columnwidth}{!}{
\begin{tabular}{|c|l|}
    \hline
    Number & \makecell[c]{
         Layer Info\\
         ($K$: Kernel Size, $C$: Output Channels, $S$: Stride)
    }  \\\hline
    $1$ &\makecell[c]{
    Conv: $K=7, C=16, S=4$\\
    Activation\\
    BatchNorm\\
    }\\\hline
    $2$ & \makecell[c]{
    Conv: $K=7, C=32, S=3$\\
    Activation\\
    BatchNorm}\\\hline
    $3$ & \makecell[c]{
    Conv: $K=5, C=64, S=2$\\
    Activation\\
    BatchNorm\\
    }\\\hline
    $4$ & \makecell[c]{
    Conv: $K=3, C=64, S=1$\\
    Activation\\
    BatchNorm\\
    }\\\hline
    $5$ & \makecell[c]{
    Conv: $K=3, C=128, S=1$\\
    Activation\\
    BatchNorm\\
    }\\\hline
    $6$ & \makecell[c]{
    Conv: $K=3, C=256, S=1$\\
    Activation\\
    BatchNorm\\
    }\\\hline
    $7$ & \makecell[c]{Adaptive Average Pool\\Flatten}\\\hline
\end{tabular}
}
\end{center}
Thus, given an input shape of $[N, 1, 2500]$ (representing a batch of $N$ 5-second samples of 1-lead ECGs sampled to 500Hz), our output shape will be $[N, 256]$.

\end{document}